\documentstyle[aps,prl,epsf,epsfig,floats,twocolumn]{revtex}
\begin{document}
\draft
\twocolumn[\hsize\textwidth\columnwidth\hsize\csname @twocolumnfalse\endcsname

\title{Quantum and thermal fluctuations in the SU(N) Heisenberg spin-glass 
model near the quantum critical point}
\author{Alberto Camjayi and Marcelo J. Rozenberg}

\address{Departamento de F\'{\i}sica, FCEN, Universidad de Buenos Aires,
Ciudad Universitaria Pab.I, (1428) Buenos Aires, Argentina.
}

\date{\today}
\maketitle
\begin{abstract}

We solve for the SU(N) Heisenberg spin-glass
in the limit of large N focusing on small $S$ and $T$.
We study the effect of quantum and thermal fluctuations
in the frequency dependent response function
and observed interesting transfers of spectral weight.
We compute the $T-$dependence of the 
order parameter and the specific heat and find an unusual $T^2$ behavior for
the latter at low temperatures in the spin-glass phase.
We find a remarkable qualitative agreement with various
experiments on the quantum frustrated magnet SrCr$_{9p}$Ga$_{12-9p}$O$_{19}$.

\end{abstract}

]

Disordered quantum magnets are fascinating systems. The understanding of the
interplay between disorder, quantum and thermal fluctuations remains among the
most challenging problems of condensed matter physics 
\cite{fisher,mvp,young,binder}. 
These three aspects are always present to some extent in experiments on
real systems, therefore a clear understanding of their interplay
is very desirable. 
In systems where disorder is relevant we usually encounter the 
phenomenology of slow
dynamics that is associated with glassy states. When quantum fluctuations
become important the phases with glassy orders can be driven to 
more conventional
phases through interesting quantum phase transitions\cite{subir}. One
example that is capturing the interest of experimentalist and theorist alike
is LiHo$_x$Y$_{1-x}$F$_4$ which is a dipolar
coupled random magnet \cite{holmio} and has been recently the focus of 
beautiful experiments \cite{brooke} where quantum fluctuations are
introduced and controlled
by means of a transverse magnetic field.  
An other example, and perhaps the archetype of frustrated quantum magnets,
is the bi-layer Kagom\'e lattice SrCr$_{9p}$Ga$_{12-9p}$O$_{19}$ (SCGO)
that only
becomes a spin-glass at the low temperature of about 5K.
This compound has been thoroughly investigated 
over the years \cite{scgo,ramirez,lee,mondelli,limot} and, 
in sharp contrast to ordinary classical
spin-glass system, exhibits some unusual remarkable features
that are associated with strong quantum fluctuations:
The magnetic fluctuation spectrum, $\chi''(\omega)$,
is found to vanish lineraly in $\omega$ at low frequencies \cite{lee}
and the specific heat is proportional to $T^2$ \cite{ramirez}. 
On the theoretical side, these 
observations have remained largely unaccounted for.

The progress in the understanding models of disordered 
quantum magnets in finite dimensions is rather slow.
In fact, a great deal of our knowledge still relies on solutions
of systems with long-ranged interactions. 
These mean-field models are 
appealing because they are mathematically more
tractable while retaining much of the physics associated with
slow dynamics.
It is worth pointing out that in many actual systems, such as
LiHo$_x$Y$_{1-x}$F$_4$ that is an insulator, 
the magnetic interactions do have power-law decay, thus
each individual spin interacts with others well beyond
their nearest neighbours\cite{holmio}.

Among the simplest
mean field models for quantum spin-glasses,
the quantum version of the Sherrington-Kirkpatrick (SK) model
received a great deal of attention.
It is a Heisenberg
model with gaussianly distributed random interactions 
between all pair of spins in the lattice.
The model was first considered by Bray and Moore \cite{bm} and they predicted 
a spin-glass 
phase at low temperature,
substantially reduced from the usual (Ising) version of the SK model.
Further progress was prevented because replica symmetry broken 
solutions were
expected at low $T$. Later, Sachdev and Ye introduced a generalization of
the model to SU(N) spins which could be studied
in the large N limit \cite{subirsum}.
They found
a very interesting spin liquid phase down to $T=0$.
In more recent work on this model, a generalized phase diagram 
as a function of $T$ and $S$ was 
obtained using a bosonic representation
\cite{gps1,gps2,kopec}.
The spin quantum number $S$ can be thought of a parameter that controls the strength
of the quantum fluctuations. For S$\to\infty$ one goes to the ``classical''
limit while for small $S$ the quantum fluctuations are strongest.
A low temperature spin-glass phase
was found for all non zero $S$ and T$_g\sim S^2$ at 
large S\cite{gps1,gps2,kopec}.
Remarkably, the spin-liquid phase was also found at very low spin S
\cite{subirsum,gps1,gps2}.
Therefore, quantum fluctuations can drive the model
through an interesting quantum critical point 
between a spin-liquid
state at S$\to$0 and a quantum spin-glass for finite S.

Recent numerical studies based
on quantum Monte Carlo \cite{qmc} and exact diagonalization \cite{ed1,ed2} 
techniques for the SU(2)
model have validated some aspects of previous investigations. 

The goal of the present work is to focus on the different roles played by
quantum and thermal fluctuations in the 
SU(N) SK model within the quantum critical regime. We 
obtain the detailed behavior of the dynamical spin susceptibility
for small $S$ and $T$, both in the paramagnetic (PM) and spin-glass (SG)
phases.
We find interesting transfers of spectral weights in the magnetic
response. We also find that the spin-glass order parameter  has
a simple temperature behavior at small $S$ and obtain the correct
specific heat at low temperatures. 
In addition, we discuss the remarkable qualitative agreement that we find
between our model solutions and the experimental results in the
SCGO compound that we mentioned above.

The model Hamiltonian is

\begin{equation}\label{hamil}
H = \frac{1}{\sqrt{{\cal N}N}} \sum_{i<j} J_{ij}
\vec{S}_{i}\cdot \vec{S}_{j},
\end{equation}
where the magnetic exchange couplings $J_{ij}$ are independent,
quenched random variables  distributed according to a Gaussian distribution
where $J$ is the variance and the unit of energy.
As already pointed out by Bray and Moore \cite{bm},
one uses the replica trick to
average over the disorder \cite{mvp} and the lattice infinite-range
model maps exactly onto a {\sl self-consistent  single
site model} with the action (in imaginary time $\tau$, with $\beta$ the
inverse temperature)  :
\begin{equation}\label{action}
S_{eff} = S_{B} - \frac{J^{2}}{2N} \int_{0}^{\beta} d\tau  d\tau'\,
Q^{ab} (\tau -\tau')
\overrightarrow{S}^{a} (\tau )\cdot \overrightarrow{S}^{b} (\tau')
\end{equation}
and the self-consistency condition
\begin{equation}\label{selfcons}
Q^{ab} (\tau -\tau') = \frac{1}{N^{2}}
<
\overrightarrow{S}^{a} (\tau )\cdot  \overrightarrow{S}^{b}(\tau')
>_{S_{eff}}
\end{equation}
where $a,b=1,\cdots,n$ denote the replica indices (the limit
$n\rightarrow 0$ has to be taken later) and $S_{B}$ is the Berry
phase of the spin \cite{subirsum}. Due to the time-dependence,
the solution of these mean-field equations remains a very
difficult problem for $N=2$, even in the paramagnetic phase \cite{qmc}.

We shall use the bosonic representation \cite{subirsum,gps1,gps2,kopec}
for the spin operators where
$S$ is
represented with Schwinger bosons $b$ by $S_{\alpha \beta}=
b^{\dagger}_{\alpha }b_{\beta } - S\delta_{\alpha \beta }$,
  with the constraint $\sum_{\alpha}
b^{\dagger}_{\alpha }b_{\alpha }= SN$  ($0\leq S$). In
the language of Young tableaux, these representations are described by
 one line of length $SN$.  They
are a natural generalization of an $SU (2)$ spin of
size $S$.

In the $N\rightarrow \infty$ limit, the mean field self-consistent
model (\ref{action}-\ref{selfcons}) reduces to
an integral equation for the Green's function of the boson
 $G_{b}^{ab} (\tau) \equiv  - \overline{<T b^{a} (\tau)
b^{\dagger b} (0)>}$ where the bar denotes the average over disorder
and the brackets the thermal average  \cite{subirsum} :
\begin{equation}\label{eq1}
(G_{b}^{-1})^{ab} (i\nu_{n}) =i\nu_{n}\delta_{ab} +
\lambda^{a}\delta_{ab} - \Sigma_{b}^{ab} (i\nu
_{n}) \\
\end{equation}
\begin{equation}\label{eq2}
\Sigma^{ab}_{b}(\tau ) =\ J^{2} \bigl( G_{b}^{ab} (\tau)\bigr)^{2}
G_{b}^{ab} (-\tau) \\
\end{equation}
\begin{equation}\label{eq3}
G_{b}^{aa} (\tau=0^{-} ) = - S
\end{equation}

The {\it local } spin susceptibility
$\chi_{loc} (\tau )=<S (\tau )S (0)>$ is given in the
large N limit by
$ \chi_{loc} (\tau )= G_{b}^{aa} (\tau)G_{b}^{aa} (-\tau )$

In the spin glass phase it is enough to perform a one step symmetry broken
solution \cite{gps1,gps2,kopec}.
Equations (\ref{eq1}-\ref{eq3}) were solved 
self-consistently on the Matsubara axis.
To obtain the imaginary part of the
$\omega$-dependent dynamical response $\chi_{loc}''(\omega)$, 
the solutions
were analytically continued
to the real axis using a method based on Pad\'e approximants \cite{note}. 
The general form of the spin susceptibility can written as 
$\chi''_{loc}(\omega) = {\rm q_{EA}}\delta(\omega) +
\chi_{reg}''(\omega)$ where ${\rm q_{EA}}$ is the spin glass order parameter.

\begin{figure}
\epsfxsize=3.in
\epsffile{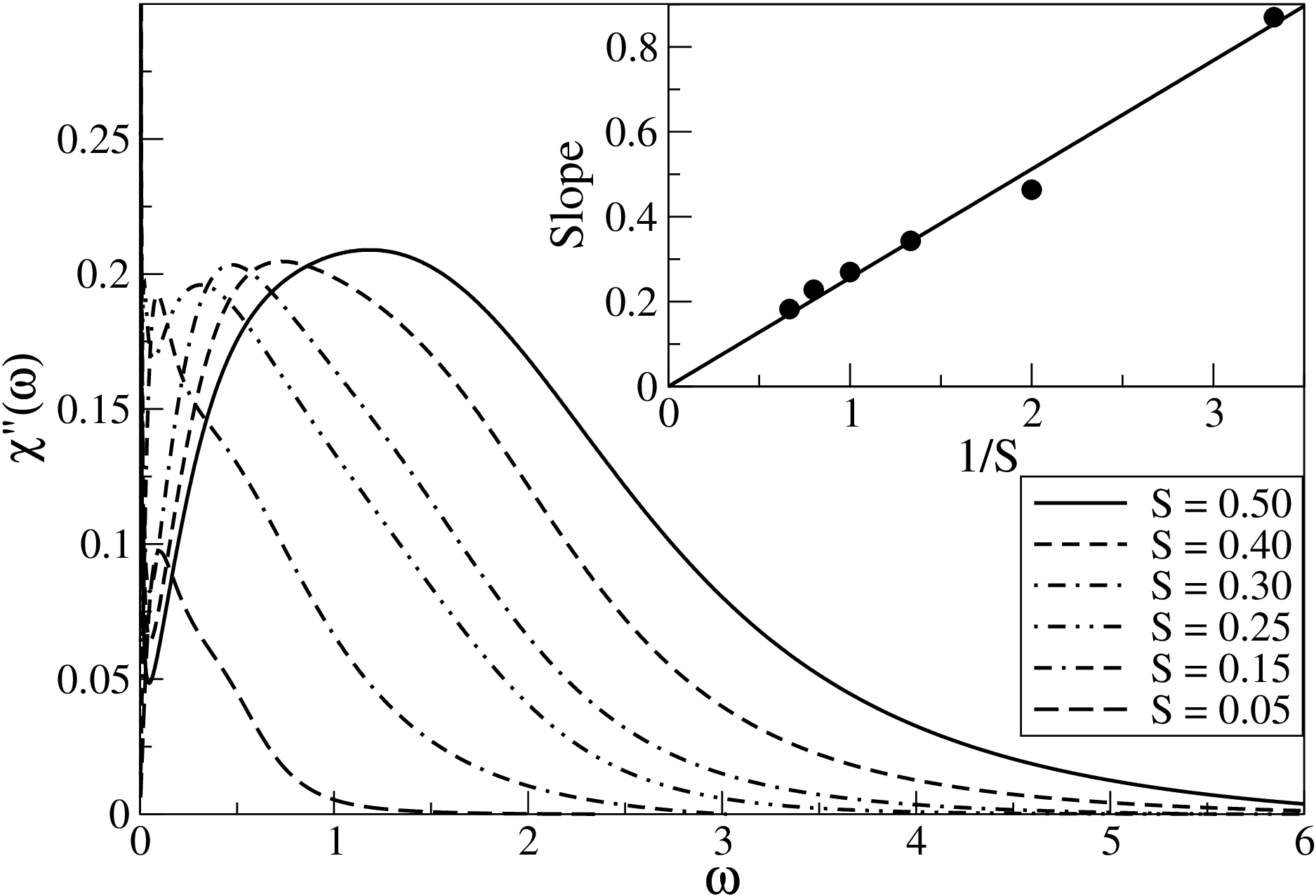}\\
\caption{
The imaginary part of the dynamical spin susceptibility $\chi''(\omega)$
as a function
of $\omega$ for various values of $S$ at $T=0.04$.
}
\label{fig1}
\end{figure}

In Fig.~\ref{fig1} we show results for $\chi_{reg}''(\omega)$ at low T=0.04 
and several values of $S$ across the PM-SG boundary.
At this $T$, the critical $S$ is found at $S \approx$0.28. 
We observe a qualitative change in the regular part of the response as
$S$ is increased. At low $S$, in the PM phase,
the susceptibility shows the finite temperature
spin liquid behavior, obeying $\chi''(\omega) \sim tanh(\omega/2T)$ for $\omega$
small. On the other hand, as $S$ increases and the system goes into the SG phase
and $\chi''_{reg}(\omega)$ opens a pseudogap.
The thermal excitations become gradually 
less important and a linear in $\omega$ behavior
shows up clearly. In fact, we find that the low
frequency behavior is proportional to $\omega/S$ (inset).
This is consistent with the large $S$
solution obtained in \cite{gps1}. 

\begin{figure}
\epsfxsize=3.in
\epsffile{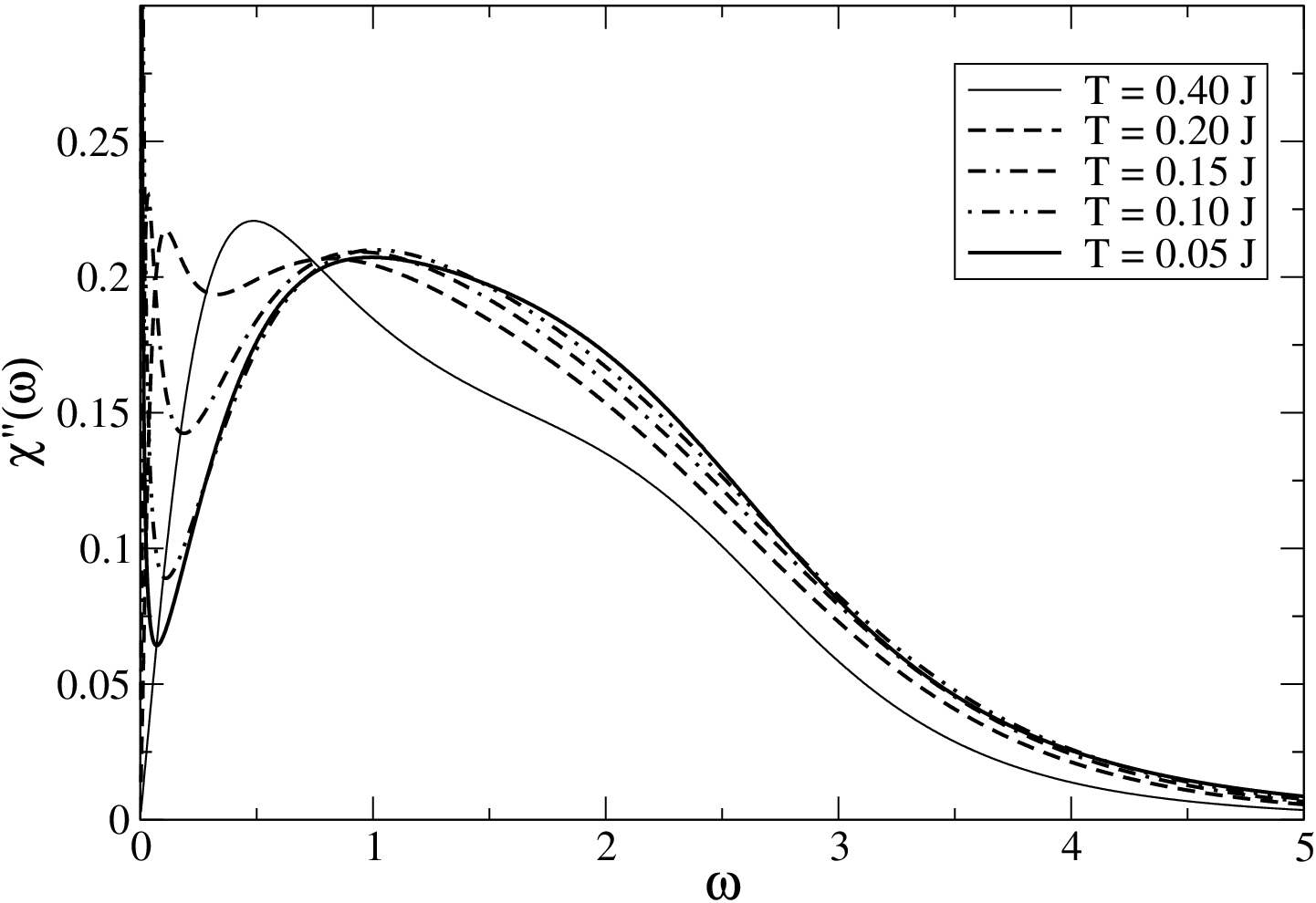}\\
  \caption{
The imaginary part of the dynamical spin susceptibility $\chi''(\omega)$
as a function
of $\omega$ for various values of $T$ at $S=0.5$. Below Tg$\approx$0.133
the systems is in the SG phase and $\chi''(\omega)$ develops a delta
function part at $\omega =0$ that is not shown for clarity.
}
\label{fig2}
\end{figure}

We now turn to the role of thermal fluctuations
for fixed $S=1/2$ that enables comparisons to numerical
results obtained in the SU(2) model \cite{qmc,ed1,ed2}.
The freezing temperature is found at $T_g\approx$0.133 
in good agreement 
with all previous estimates \cite{bm,qmc,ed1,ed2}.
We start at
low $T=0.05$ well in the spin-glass phase. In Fig.~\ref{fig2} we
show the susceptibility
with clean pseudogap $\sim \omega$ behavior at low $\omega$ 
(the $\delta(\omega)$ is not shown for clarity).
As $T$ is increased, one observes that excitations gradually
fill the pseudogap
with a narrow low frequency feature that peaks
at $\omega \sim {\cal O}(T)$.
These excitations come from the gradual melting of the frozen spins, i.e.,
from the $\delta$-function part. 
Another interesting effect that one observes is that spectral weight 
from {\em high} frequencies of order $J$ is transferred down
to the pseudogap. The interpretation of this is that when spins are
frozen in the spin glass state, they still  
have a fast motion of precession around the axis of their local frozen
field. That motion originates a contribution to the susceptibility
at $\omega \sim {\cal O}(J)$ with a strength proportional to the 
frozen fraction,
i.e., to the order parameter q$_{EA}$. As $T$ increases, the spins (and thus
their local field) melt, so the contribution from the 
motion of precession gradually decreases and merges with the
excitations of order $T$ that now fill the pseudogap.
We may point out that
this behavior is qualitatively similar to the results obtained
from exact diagonalization of small SU(2) clusters \cite{ed2}.

As $T$ is further increased one enters the PM phase and the melted peak 
fully merges with the higher frequency part of $\chi''(\omega)$ 
and there is no more a clear separation of energy scales. In this
quantum disordered regime the
low frequency behaviour of $\chi''(\omega)$ is $\propto tanh(\omega/2T)$
as in the spin-liquid state \cite{subirsum}.

Thus we have seen that the regular part of the response begins
at low $T$ in the SG phase with a clean
and linear in $\omega$ pseudo-gap,
then above $T_g$ the gap becomes thermally filled down to very low 
frequencies,
and finally when $T$ is well above $T_g$ the pseudo-gap clears up 
again displaying
once more a linear in $\omega$ behavior. We find remarkable that
this unusual evolution is 
qualitatively identical to that reported from neutron experiments in
SCGO (cf. Fig.3 of Ref.\cite{lee}). 
Moreover, the neutrons have also revealed that the spatial correlations
are extremely short ($\sim 2.5\AA$) 
which may render additional justification to the
relevance of the present mean-field theory results.

\begin{figure}
\epsfxsize=3.in
\epsffile{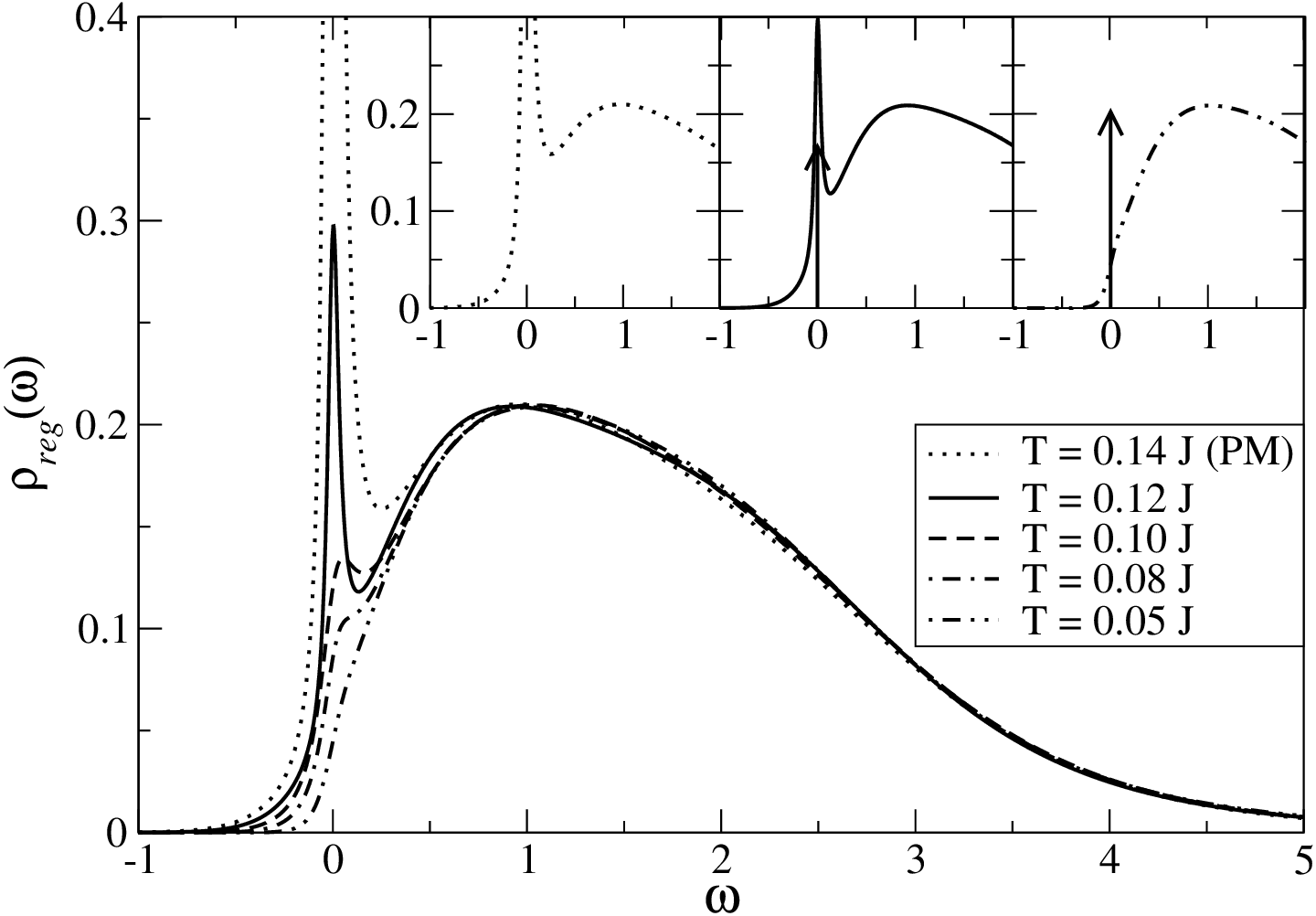}\\
  \caption{
The regular part of spectral density function $\rho(\omega)$ 
as a function
of $\omega$ for various values of $T$ at $S=1/2$. Below $T_g\approx$0.133
the systems is in the SG phase. The inset shows $\rho(\omega)$ in the
PM phase (left), entering the SG phase (center) and at low $T$ within
the SG (right). The height of the arrow indicates the spectral intensity
of the delta part at the origin.
}
\label{fig3}
\end{figure}

In order to better appreciate the evolution of the transfers of spectral
weight at low frequencies and for small $T$,
it is useful to consider the spectral
density, defined by 
$\rho(\omega) = \chi''_{loc}(\omega)/ (e^{-\beta\omega}-1)$ 
that obeys the sum-rule $\int \rho_{reg}(\omega)d\omega + q_{EA} = S(S+1)$.
The spectral density at different temperatures is shown in Fig.~\ref{fig3}
where the intensity of the delta function part is denoted by the height
of the arrow (see inset) and can be thought as the fraction
of frozen spins. In the main panel we show the regular part
of the spectral density that has a broad background contribution that remains almost temperature independent. Most of the $T-$dependence
occurs at the low frequencies
where a rather narrow peak is present. At the higher temperature, in the
PM phase there is no delta function contribution, however the
large peak at small $\omega$
indicates that a portion of the degrees of freedom actually got slowed down (left panel 
of inset).  As $T$ is lowered, the system enters the SG phase and the 
peak becomes narrower and losses weight. A $\delta$-function contribution 
thus emerges as some of the slow spins
become frozen (central panel of inset). When $T$ is further lowered
towards $T=0$ we observe how the resonance losses all its
weight that gets transferred to the $\delta$ part (right panel of inset). 
The strong
quantum fluctuations are responsible for the large remanent background 
spectral density that corresponds to a large fraction of spins
remaining disordered. 

\begin{figure}
\epsfxsize=3.in
\epsffile{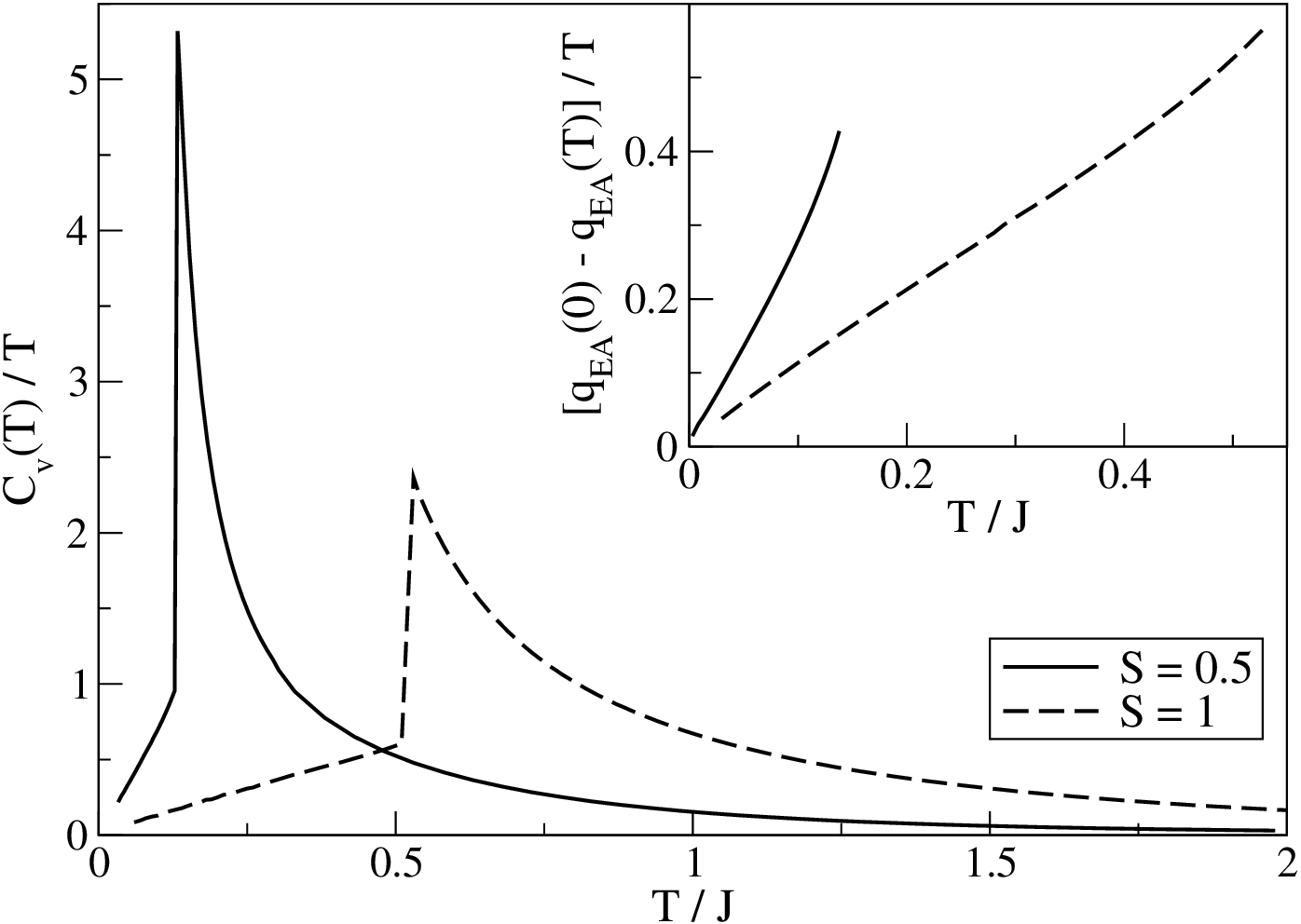}\\
  \caption{
The specific heat $C_v(T)/T$ for $S=1/2$ and $1$. Inset:
The SG order parameter $[q_{EA}(T=0)-q_{EA}(T)]/T$ 
as a function of $T$ for the same values
of $S$. 
}
\label{fig4}
\end{figure}

The results for the spin-glass order parameter q$_{EA}(T)$ are shown 
in the inset of Fig.~\ref{fig4}. The behavior at moderate and
large values of $S$ 
was previously investigated in
Ref.\cite{gps2}.
We focus here in the small $S$ and $T$ regime and find that 
the order parameters obeys the simple form 
\begin{equation}
q_{EA}(T)= q_{EA}(T=0) - \alpha T^2
\label{qea}
\end{equation}
with $\alpha$ a constant,
which is similar to the solution of the SK 
model\cite{fisher,sk}.
We also find that the value of $q_{EA}$ at $T=0$ and its jump at $T_g$(S) are
strongly reduced by quantum fluctuations
when $S\to 0$. In fact, the later vanishes faster than $S^3$
in contrast to the quadratic dependence at large $S$ \cite{gps2}. 
It is interesting to note the good agreement 
between $q_{EA}(T=0) \approx 0.20$ for $S=1/2$ with the corresponding
estimate q$_{EA}(T=0) \approx 0.18$ obtained in an exact diagonalization
study of the SU(2) model \cite{ed2}.
Possibly more significant is to mention that a frozen moment
of only 20$\%$ of the total possible elastic value was observed in
neutron scattering on SCG0 \cite{lee,mondelli}.

Finally we turn to the specific heat $C_v$. 
In earlier work \cite{gps1}, it was argued
that the specific heat is linear in $T$ at low temperatures within the SG phase.
We calculated the specific heat in the whole range of $S$ values,
with particular care in the small $S$ regime.
In contrast to previous results\cite{note2}, we find that the behavior
within the SG phase at low $T$ has the form $C_v(T) \propto T^2$
as is shown in Fig.~\ref{fig4}.
This result can be justified by the following argument:
The energy of the system is given by,
\begin{equation}
E(T)= -\frac{J^2}{2}\int^\beta_0 d\tau Q^{ab}(\tau) Q^{ab}(-\tau).
\label{energy}
\end{equation}
Writting this expression in the SG phase in terms of the one-step
replica symmetry broken solution \cite{gps1} and
inserting the functional form of the order parameter (\ref{qea}),
one finds that the lowest order terms in the $T-$expansion
of $E(T)$ are proportional to $T^{-1}$, $T^1$ and $T^3$.
It is not hard to check that the coefficient of the divergent
$T^{-1}$ term exactly cancels. On the other hand, the linear
term should also vanish as otherwise 
the groundstate would have extensive remanent
entropy. Therefore, one may expect that the
first non vanishing contribution to the energy at low $T$ is proportional
to $T^3$, which implies a quadratic behavior for $C_v(T)$ as is in fact
borne out of our calculations. While ordinary spin-glasses usually show a
linear in $T$ specific heat, we find remarkable that the unusual
$T^2$ behavior that we obtain was indeed observed 
among the early investigations of the spin-glass state
of the SCGO compound \cite{ramirez} that pointed to the important
role that quantum fluctuations might play in that compound.

To conclude, we have solved for the behavior of the 
Heisenberg SU(N) spin-glass in the limit of large N and 
investigated in detail the parameter region of small $S$ and $T$
close to the quantum critical point of the model.
We observed the qualitatively different role played
by quantum and thermal fluctuations through
their effect on the transfers of spectral weight in the dynamical
response functions. 
We obtained the functional form of the 
order parameter and present an argument to support the finding of an
unusual quadratic behavior of the specific heat at low temperature
within the spin glass phase.
We find a very interesting qualitative agreement with various experimental
findings in the SCGO compound one of the most thoroughly investigated
quantum spin-glass system. Extensions of our work to incorporate a more
realistic geometric structure might be interesting routes for future research.

We acknowledge support of
Fundaci\'on Antorchas, CONICET (PID $N^o4547/96$), ANPCYT
(PMT-PICT1855) and ECOS-SeCyT. MJR acknowledges also the hospitality
at the KITP of the UCSB where part of this work was completed.


\begin{thebibliography}{10}

\bibitem{fisher}  K.H.Fischer and J.A.Hertz, Spin Glasses, Cambridge
University Press, Cambridge, England (1991).

\bibitem{mvp} M. M\'ezard, G. Parisi and M. Virasoro, {\it Spin Glass
Theory and Beyond} (World Scientific, Singapore 1987).

\bibitem{young}  H.Rieger und A.P.Young, {\it Quantum Spin Glasses},
Lecture Notes in Physics {\bf 492} ''Complex Behavior of Glassy Systems'',
p. 254, ed. J.M. Rubi and C. Perez-Vicente (Springer Verlag,
Berlin-Heidelberg-New$\,$York, 1997).

\bibitem{binder}  K.Binder and A.P.Young, Rev.Mod.Phys. {\bf 58}, 801
(1986).

\bibitem{subir} S. Sachdev. Quantum Phase Transitions, 
Cambridge University Press,
Cambridge, England (1999).

\bibitem{holmio}  D.H.Reich {\it et al.}, Phys. Rev. B {\bf 42}, 4631
(1990).  W. Wu  {\it et al.}, Phys.
Rev. Lett. {\bf 67}, 2076 (1991). W. Wu  {\it et al.}, Phys. Rev. Lett. {\bf 71}
, 1919 (1993).

\bibitem{brooke}  J.Brooke, D.Bitko, T.F.Rosenbaum and G.Aeppli, Science
{\bf 284}, 779 (1999).

%
%

\bibitem{scgo} X. Obradors {\em et al}, Solid State Commun. {\bf 65},
189 (1990).

\bibitem{ramirez} A. Ramirez {\em et al}, Phys. Rev. B {\bf 45}, 2505 (1992).

\bibitem{lee} S.-H. Lee {\em et al}, Europhys. Lett. {\bf 35}, 127 (1996).

\bibitem{mondelli} C. Mondelli {\em et al}, Physics B {\bf 266} 104 (1999).

\bibitem{limot} L. Limot {\em et al}, Phys. Rev. B {\bf 65}, 144447 (2002)
and references therein.

\bibitem{bm}  A.J.Bray and M.A.Moore, J. Phys. C {\bf 13}, L655 (1980).

\bibitem{subirsum}  S. Sachdev and J. Ye, Phys. Rev. Lett. {\bf 70}, 3339
(1993).

\bibitem{gps1}   A. Georges, O.Parcollet and S. Sachdev,
Phys. Rev. Lett. 85, 840 (2000).

\bibitem{gps2}   A. Georges, O.Parcollet and S. Sachdev,
Phys. Rev. B {\bf 63}, 134406 (2001).

\bibitem{kopec} T.K. Kope\'c, Phys. Rev. B {\bf 52}, 9590 (1995).


\bibitem{qmc}  D.R.Grempel and M.J.Rozenberg, Phys. Rev. Lett.
{\bf 80}, 389 (1998).

\bibitem{ed1}  L.Arrachea and M.J.Rozenberg,
Phys. Rev. Lett. {\bf 86}, 5172 (2001).

\bibitem{ed2}  L.Arrachea and M.J.Rozenberg,
Phys. Rev. B {\bf 65}, 224430 (2002).


\bibitem{note}
We checked the reliability of the analytic continuation 
for different choices of number of frequency points.

\bibitem{sk}  D.Sherrington and S.Kirkpatrick, Phys. Rev. Lett. {\bf 35},
1972 (1975).

\bibitem{note2}
To obtain accurate results we had to use upto
about 2 million Matsubara frequency points.


\end{thebibliography}
\end{document}